\documentclass[runningheads]{llncs}
\usepackage{url}
\usepackage{lipsum}
\usepackage{graphicx}
\usepackage{fancyhdr}
\usepackage{amsmath}

\def\rk {R_K}
\def\rkst {R_{K^*}}
\def\Babar{{\mbox{\slshape B\kern-0.1em{\it A}\kern-0.1em B\kern-0.1em{\it A\kern-0.2em R}}}}
\def\beq{\begin{equation}}
\def\eeq{\end{equation}}
\newcommand{\be}{\begin{equation}}
\newcommand{\ee}{\end{equation}}
\newcommand{\ba}{\begin{eqnarray}}
\newcommand{\ea}{\end{eqnarray}}

\def\g {\gamma}
\begin{document}

\title{Flavor violation at LHC in events with two opposite sign leptons and a b-jet}
\titlerunning{flavor}
\author{Nilanjana Kumar \inst{1}}
\institute{$^1$ Department of Physics and Astrophysics, University of Delhi, Delhi 110007, India\\
\email{nilanjana.kumar@gmail.com} 
}
\maketitle              
\vspace{-2mm}
\begin{abstract}
Hints of flavor violation at both charged current and neutral current 
decays have been observed in experiments such as LHCb, Belle, and \Babar. 
The anomalies in the result can be addressed in the effective field theory 
(EFT) framework. 
The effective operators predict different 
beyond standard model (BSM) signatures and the four point interaction vertices can be probed 
at Large Hadron Collider (LHC). In this context,  
the discovery projection of two opposite sign leptons and a b-jet signature is 
studied in this paper at 13 TeV LHC. 
\keywords{Flavor violation, Opposite-sign leptons}
\end{abstract}
\vspace{-4mm}
\section{Motivation}
\vspace{-1mm}
Recent experimental measurements at LHCb, Belle, and \Babar~present 
deviations in the SM prediction of B-meson decays and hints 
towards Lepton Flavor Violation(LFV). 
LFV has been observed in charged current decay at 
tree level, $b \rightarrow c \ell \nu$. Taking into account
$R(D)$ and $R(D^{*})$ measurements by \Babar~\cite{Lees:2013uzd},
Belle \cite{Huschle:2015rga} and LHCb \cite{Aaij:2015yra} 
and their correlations, the difference between SM and the 
data is nearly $3.8\sigma$ \cite{Amhis:2014hma,hflav}. 
The anomaly in $B_c \to J/\psi \ell \nu$ measurement is at the $2\sigma$ level. 
Whereas, the neutral current transitions,
namely $b \to s \ell^+ \ell^-$ shows an opposite effect in 
the measurements of $\rk$ and $\rkst$. The recent results by 
LHCb collaboration~\cite{CERN-EP-2019-043} and Belle~\cite{Abdesselam:2019wac},
reflect that the data is more consistent with the SM. 
A deviation is also seen in $B_s\to \phi\mu\mu$~\cite{Aaij:2015esa},  
that suggests that the discrepancies in $\rk$ and $\rkst$ have been 
caused by a diminution of the $b \to s
\mu^+ \mu^-$ channel, rather than an enhancement in $b \to s e^+ e^-$.

To address these anomalies one can choose the models with 
leptoquarks~\cite{Becirevic:2016yqi} or $Z'$~\cite{Langacker:2008yv}. 
Another way of addressing these anomalies would be 
to consider an effective field theory (EFT) description with a 
set of Wilson coefficients. It is possible to construct such a theory 
with a few unknown parameters, if symmetry relations exist between the 
Wilson Coefficients. As shown 
recently in~\cite{Bhattacharya:2019eji,Choudhury:2017qyt,Choudhury:2017ijp},  
with a minimal set of new physics (NP) operators, accompanied 
by a single lepton mixing angle, it is possible to address 
the flavor observables. The parameters of these models 
can be determined phenomenologically and if the scale of the 
new physics is a few TeVs, this leads to interesting
collider signatures at LHC.
\section{Theoretical Framework}
The Hamiltonian for the new physics can be expressed in terms of two
operators involving left handed 2nd and 3rd generation 
quark doublets $Q_{2L}$, $Q_{3L}$, 3rd generation lepton doublet 
$L_{3L}$ and right handed singlet $\tau_{R}$ as defined in 
Ref.~\cite{Choudhury:2017qyt,Choudhury:2017ijp}, where 
The terms that we are interested in are 
$(\bar Q_{2L} \g^\mu Q_{3L})_3 \, (\bar L_{3L} \g^\mu L_{3L})_3$
and 
$(\bar Q_{2L} \g^\mu Q_{3L})_1 \, (\bar \tau_R \g^\mu \tau_R)$, 
with coefficients $3A_{1}/4$ and $A_5$, following the literature 
Ref.~\cite{Choudhury:2017qyt,Choudhury:2017ijp}.
$A_{i}$ are real unknown coefficients with dimension TeV$^{-2}$. 
The subscripts `3' and `1' represent the
$SU(2)_L$ triplet and singlet currents respectively.
The flavor eigenstates can be expressed in terms of mass eigenstates by 
a field rotation, 
\begin{center}
$\tau = cos\theta(\tau') + sin\theta(\mu')$ \\
$\nu_\tau = cos\theta(\nu_\tau') + sin\theta(\nu'_\mu)$\\
\end{center}
As a result of the mixing, the coupling with the 
second generation of leptons are induced. The magnitude of this 
mixing is found to be small ($\sim0.02)$ ~\cite{Choudhury:2017qyt,Choudhury:2017ijp}.
Also, for all class of models the best fit values obtained can be approximated as   
$A_1\sim$3.8, $A_5\sim$2.3.

The flavor violating process, generated by these operators 
are listed in Table~\ref{tab:1}. 
For a process (a,b) $\rightarrow $(c,d) 
with coefficient X in the operator, we can write the four point coupling ($\lambda^2$)
in the mass basis as,
\begin{center}
$\frac{\lambda^{*}_{a,b}\lambda_{c,d}}{2M^2} (a,b)(c,d) = \epsilon^{abcd} \frac{4G_F}{\sqrt 2}(a,b)(c,d)$,\\
$\lambda^2\sim\lambda^{*}_{a,b} \lambda_{c,d}\sim 2 M^2 X$,
\end{center}
where, M is the mass of the integrated-out field, and $\lambda$'s are the dimensionless 
coupling. From perturbativity the 
bound on $\lambda$ is $\lambda^2/(4\pi)^2 \leq 1$.
\begin{table}[tpb]
  \begin{center}
    \caption{\label{tab:1}Operators and their effective couplings. For the notation, 
see Ref.~\cite{Choudhury:2017qyt}.}
\begin{tabular}{|c|c|c|c|}
\hline
Flavor basis & Mass basis& $\lambda^2$ \\
\hline
$(3A_1/4)(s,b)(\tau \tau)$&$(3A_1/4)cos^2\theta(s,b)(\tau' \tau')_L$ & $2M^2 (3A_1/4)cos^2\theta$\\\
-&$(3A_1/4)sin^2\theta (s,b)(\mu' \mu')_L $& $2M^2 (3A_1/4)sin^2\theta$ \\
-&$(3A_1/4)sin2\theta (s,b)(\mu' \tau')_L$& $2M^2 (3A_1/4)sin2\theta$ \\
\hline
$A_5(s,b)(\tau \tau)$&$A_5 cos^2\theta(s,b)(\tau' \tau')_R$& $2M^2 A_5 cos^2\theta$ \\
-                    &$A_5 sin^2\theta(s,b)(\mu' \mu')_R $ & $2M^2 A_5 sin^2\theta$ \\
-                    &$A_5 sin2\theta(s,b)(\mu' \tau')_R $ & $2M^2 A_5 sin2\theta$ \\
\hline
\end{tabular}
\end{center}
\end{table}
\section{Results}
As can be seen from Table~\ref{tab:1}, there are three possible signatures:\\
\begin{itemize}
\item($\mu^\pm \mu^\mp$) + $b$-jet,
\item ($\mu^\pm \tau^\mp$) + $b$-jet and 
\item ($\tau^\pm \tau^\mp$) + $b$-jet.
\end{itemize} 
These processes can be generated at 13 TeV 
LHC via $g$-$g$ and $g$-$s$ fusion in $p$-$p$ collision, with the major contribution coming from $g$-$s$ 
fusion. Now as the coupling ($\lambda^2$) is a function of $A_1$ and $A_5$, 
the cross section also varies with these parameter. We kept the value of 
$A_1$ fixed at the best fit 3.8 and varied $A_5$. 
The range of values of the parameters are chosen such that the  
95\% C.L. upper bound of $Br (B_s \to \tau^\pm
\mu^\mp) < 4.2 \times 10^{-5}$~\cite{Aaij:2019okb} is satisfied.

\begin{figure}[!h]
\begin{center}
\includegraphics[width=5cm,height=4.4cm]{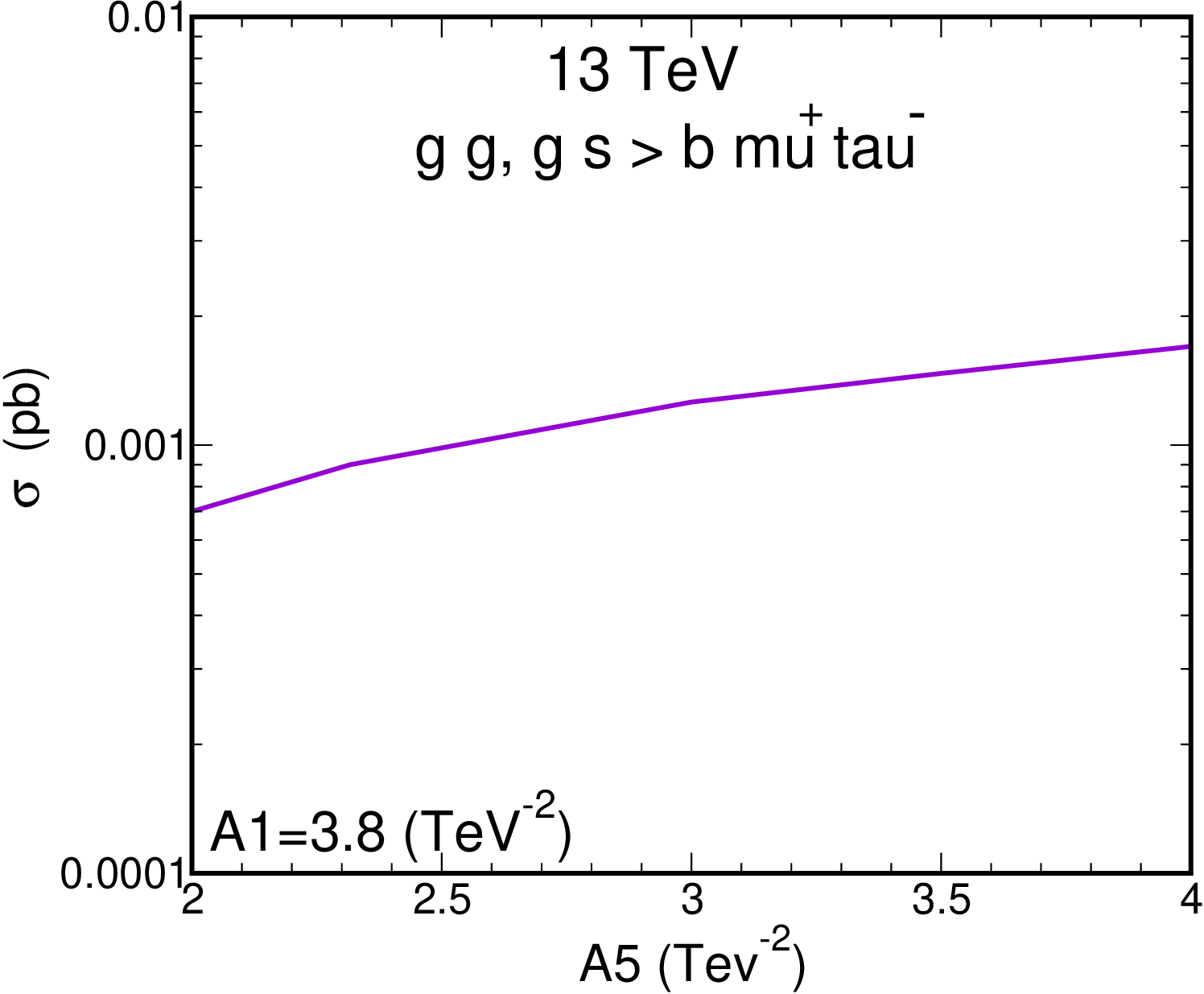}
~~~\includegraphics[width=5cm,height=4.4cm]{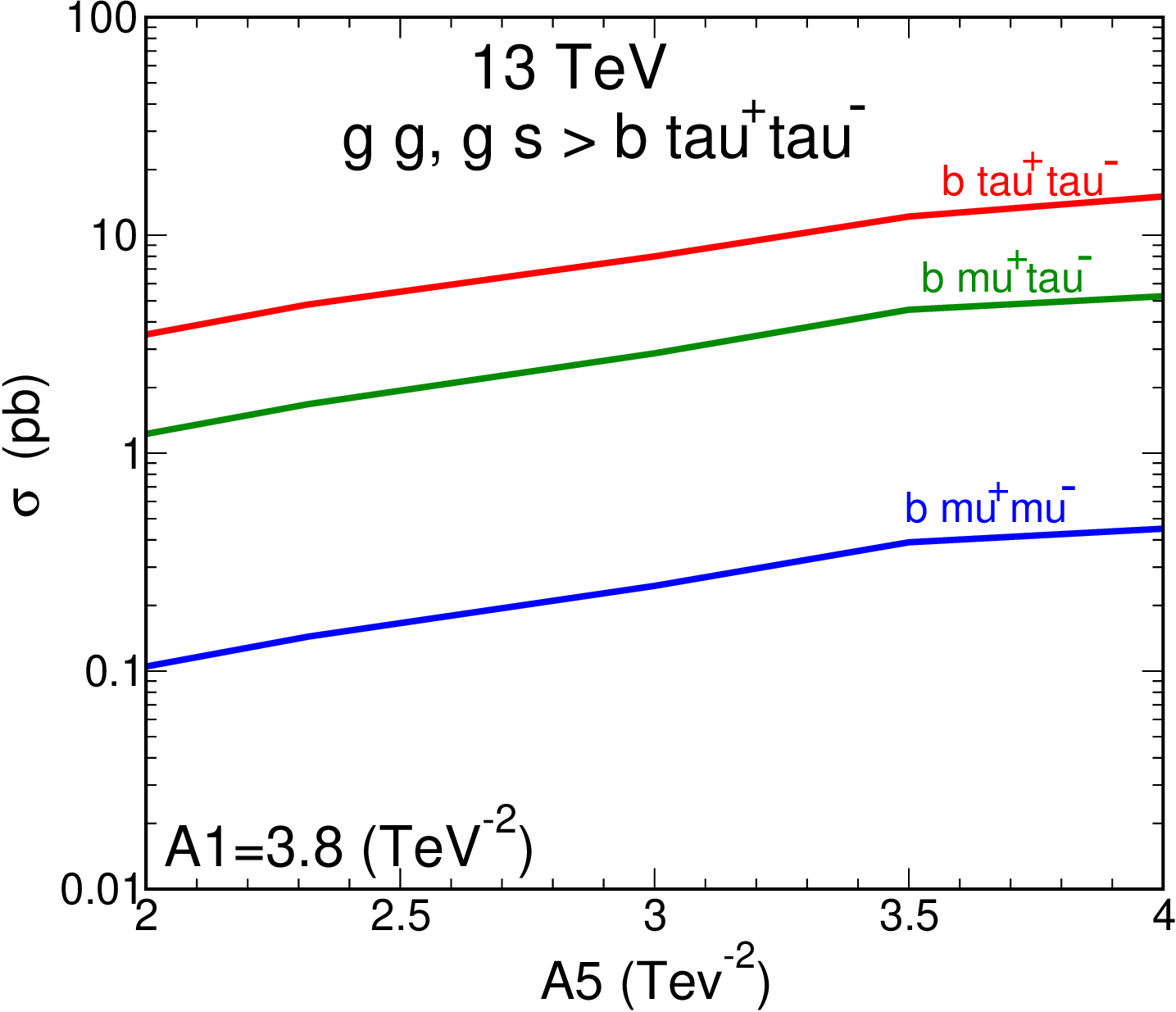}
\end{center}
\caption{\label{fig:1}{\it (L) The total production of 
one $\mu$ and one $\tau$ in association with one $b$-jet in $g$-$g$ fusion 
and  $g$- $s$ fusion. (R) The total production of 
two $\tau$,s in association with one $b$-jet in $g$-$g$ fusion 
and  $g$- $s$ fusion (red), also one $\mu$ and one $\tau$ + $b$-jet (green) and 
two $\mu$'s+$b$-jet (blue) when tau (one or both respectively) decays leptonically.}}
\end{figure}

The cross section of 
($\mu^\pm \mu^\mp$) and a $b$-jet 
is very small, because their coupling is suppressed by ($\sin^2\theta$) and hence 
we neglect it in this study. 
The cross section of ($\mu^\pm \tau^\mp$) and a $b$-jet will be suppressed 
by $\sin2\theta$ and hence will be comparatively larger than the previous one,
shown in Fig~\ref{fig:1} (left). 
The cross section of ($\tau^\pm \tau^\mp$) and a $b$-jet is 
relatively very high as shown in Fig~\ref{fig:1}(right) with red line. 
The tau can also decay leptonically to a  
muon with branching ratio 0.174, enabling the final states with 
($\mu^\pm \tau^\mp$) and a $b$-jet and ($\mu^\pm \mu^\mp$) and a $b$-jet, 
as shown in Fig~\ref{fig:1} (right) by green and blue lines respectively.

\begin{figure}[!h]
\begin{center}
\includegraphics[width=5.6cm,height=5cm]{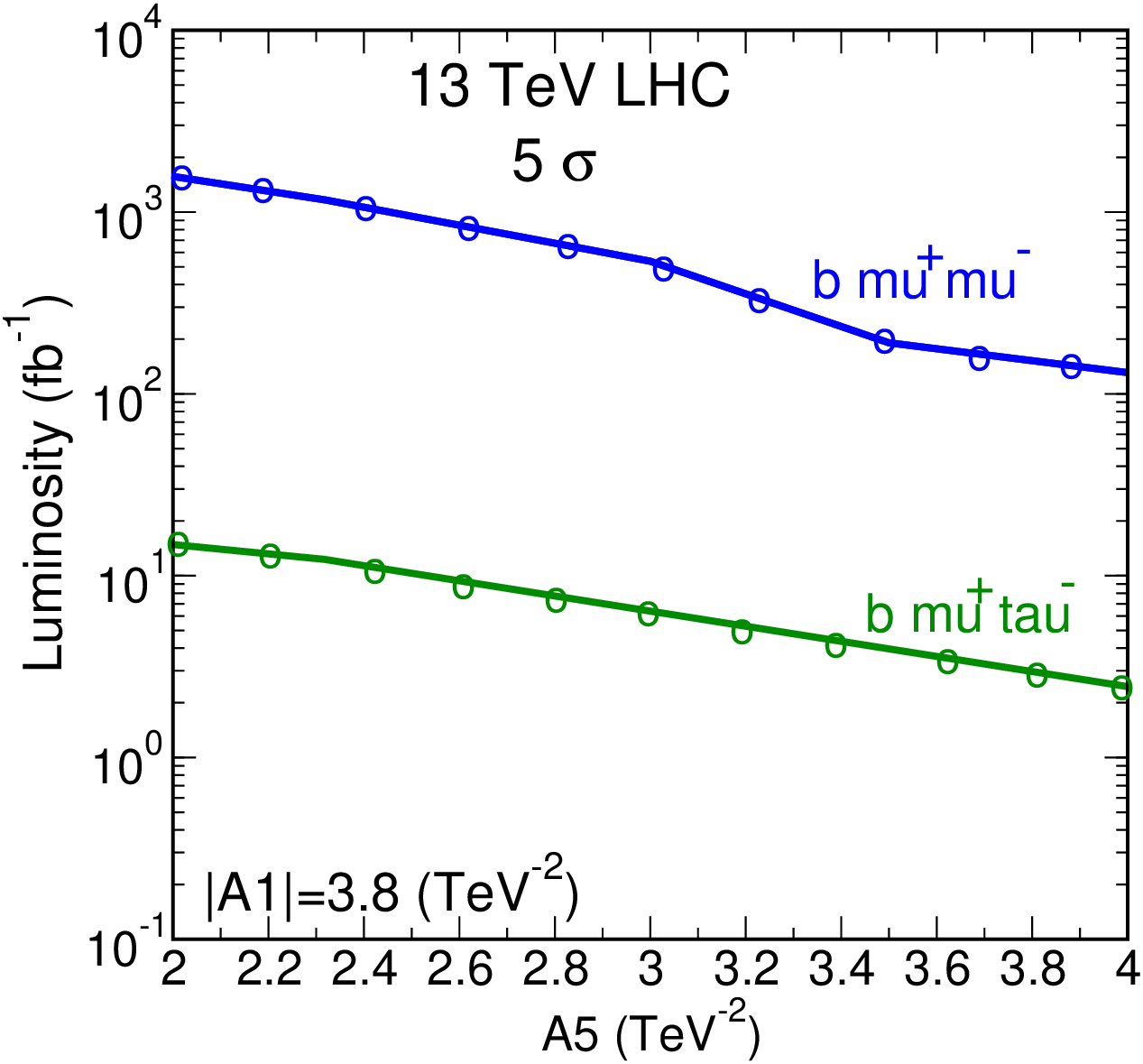}
\end{center}
\caption{\label{fig:2}{\it The 5$\sigma$ discovery projection at 13 TeV LHC in 
($\mu^\pm \mu^\mp$) + $b$-jet and ($\mu^\pm \tau^\mp$) + $b$-jet channel as 
a function of the integrated luminosity and model parameter 
$A_5$ with the assumption of 25\% uncertainty in the background events.}}
\end{figure}

The major SM backgrounds for these channels are $t \overline{t}$, 
single top ($Wt$), diboson ($W^+W^-$, $WZ$ and $ZZ$), $W$+jets, $WW$+jets, $Z/\gamma$+jets. 
We found that the background for the signal with opposite sign same flavor states 
($\mu^\pm \mu^\mp$)or ($\tau^\pm \tau^\mp$) and a $b$-jet 
are larger than the opposite sign opposite flavor state ($\mu^\pm \tau^\mp$). Also, 
the signal ($\tau^\pm \tau^\mp$)and a $b$-jet will suffer from 
tau tagging efficiency at LHC as the both the tau decay hadronically.
Hence we study two channels, ($\mu^\pm \mu^\mp$)+$b$-jet and ($\mu^\pm \tau^\mp$)+$b$-jet. 
In Fig~\ref{fig:2} we have shown the discovery projection of  
these two channels 
as a function of the integrated luminosity at LHC. We have followed the 
search strategy as mentioned in~\cite{Choudhury:2019ucz,Afik:2018nlr}.
Fig~\ref{fig:2} shows that the ($\mu^\pm \mu^\mp$)+$b$-jet channel requires much larger 
luminosity than ($\mu^\pm \tau^\mp$)+$b$-jet channel for 5$\sigma$ discovery significance.

\section{Conclusion}
Recently observed anomalies in the decays of B-mesons hints towards 
new physics interaction which involves a b-quark, a s-quark and a 
pair of opposite sign leptons. The four point interactions among them can be probed 
at LHC p-p collision via the direct production of b-quark and two opposite 
sign leptons. The opposite sign lepton pair have either same or opposite flavor. 
In this study, 5$\sigma$ discovery potential of ($\mu^\pm \mu^\mp$) + $b$-jet and 
($\mu^\pm \tau^\mp$) + $b$-jet channels are discussed as a function of 
the effective field theory model parameters. Overall, these channels have 
a very good detection prospect even with the currently collected data at LHC and 
a limit on the model parameter space can be set with the 13 TeV LHC data. 



\vspace{-2mm}
\begin{thebibliography}{6}
\vspace{-2mm}

\bibitem{Lees:2013uzd}
{\scshape BaBar} collaboration, J.~P. Lees et~al., 
{\emph{Phys. Rev.} {\bf D88} (2013) 072012}, 

\bibitem{Huschle:2015rga}
{\scshape Belle} collaboration, M.~Huschle et~al., 
{\emph{Phys.Rev.} {\bf D92} (2015) 072014}, 

\bibitem{Aaij:2015yra}
{\scshape LHCb} collaboration, R.~Aaij et~al., 
{\emph{Phys. Rev. Lett.} {\bf 115} (2015) 111803}, 

\bibitem{Amhis:2014hma}
{\scshape Heavy Flavor Averaging Group (HFAG)} collaboration, Y.~Amhis et~al.,
arXiv: 1412.7515 [hep-ph].

\bibitem{hflav}
``{https://hflav-eos.web.cern.ch/hflav-eos/semi/summer18/RDRDs.html}.''.

\bibitem{CERN-EP-2019-043}
{\scshape LHCb Collaboration} collaboration, 
Tech.  Rep. CERN-EP-2019-043. LHCB-PAPER-2019-009, CERN, Geneva, Mar, 2019.

\bibitem{Abdesselam:2019wac}
{\scshape Belle} collaboration, A.~Abdesselam et~al., 

\bibitem{Aaij:2015esa}
{\scshape LHCb} collaboration, R.~Aaij et~al., 
{\emph{JHEP} {\bf 09} (2015)  179}, 

\bibitem{Becirevic:2016yqi}
D.~Bečirević, S.~Fajfer, N.~Košnik and O.~Sumensari, 
{\emph{Phys. Rev.} {\bf D94} (2016) 115021}, 

\bibitem{Langacker:2008yv}
P.~Langacker, 
{\emph{Rev. Mod. Phys.}{\bf 81} (2009) 1199--1228}

\bibitem{Choudhury:2017qyt}
D.~Choudhury, A.~Kundu, R.~Mandal and R.~Sinha, 
{\emph{Phys. Rev. Lett.} {\bf 119} (2017) 151801},

\bibitem{Choudhury:2017ijp}
D.~Choudhury, A.~Kundu, R.~Mandal and R.~Sinha, 
{\emph{Nucl. Phys.}{\bf B933} (2018) 433--453}, 

\bibitem{Bhattacharya:2019eji}
S.~Bhattacharya, A.~Biswas, Z.~Calcuttawala and S.~K. Patra, 
arXiv:1902.02796 [hep-ph].

\bibitem{Aaij:2019okb}
{\scshape LHCb} collaboration, R.~Aaij et~al., \emph{{Search for the
  lepton-flavour-violating decays $B^{0}_{s}\to\tau^{\pm}\mu^{\mp}$ and
  $B^{0}\to\tau^{\pm}\mu^{\mp}$}},
  arxXiv:1905.0661 [hep-ph].

\bibitem{Choudhury:2019ucz} 
  D.~Choudhury, N.~Kumar and A.~Kundu,
  arXiv:1905.07982 [hep-ph].

\bibitem{Afik:2018nlr}
Y.~Afik, J.~Cohen, E.~Gozani, E.~Kajomovitz and Y.~Rozen, \emph{{Establishing a
  Search for $b \rightarrow s \ell^{+} \ell^{-}$ Anomalies at the LHC}},
{\emph{JHEP} {\bf 08} (2018) 056}.

\end{thebibliography}
\end{document}